\documentclass[12pt]{article}
\usepackage[utf8]{inputenc}
\usepackage{amsmath,graphicx}
\usepackage[style=vancouver]{biblatex}
\renewcommand{\vec}[1]{\boldsymbol{#1}}
\usepackage{fullpage}
\usepackage{lineno}
\usepackage{xcolor}
\usepackage{pdflscape}
\usepackage{hyperref}
\usepackage{authblk}

\title{\sc{Parameter identifiability, parameter estimation and model prediction for differential equation models}}

\author[1,2]{Matthew J. Simpson}
\author[3]{Ruth E. Baker}
\affil[1]{School of Mathematical Sciences, Queensland University of Technology (QUT), Brisbane, Australia.}
\affil[2]{ARC Centre of Excellence for the Mathematical Analysis of Cellular Systems, QUT, Brisbane,  Australia.}
\affil[3]{Wolfson Centre for Mathematical Biology, Mathematical Institute, University of Oxford, Oxford, OX2 6GG, United Kingdom}

\date{}

\begin{document}


\maketitle

\begin{abstract}
Interpreting data with mathematical models is an important aspect of real-world industrial and applied mathematical modeling.  Often we are interested to understand the extent to which a particular set of data informs and constrains model parameters.  This question is closely related to the concept of parameter identifiability, and in this article we present a series of computational exercises to introduce tools that can be used to assess parameter identifiability, estimate parameters and generate model predictions.  Taking a likelihood-based approach, we show that very similar ideas and algorithms can be used to deal with a range of different mathematical modeling frameworks.  The exercises and results presented in this article are supported by a suite of open access codes that can be accessed on \href{https://github.com/ProfMJSimpson/IdentifiabilityTutorial}{GitHub}.
\end{abstract}

\newpage 
\section{Introduction} \label{sec:intro}
Parameter estimation is a critical step in real-world applications of mathematical models that enables scientific discovery and decision making across a broad range of applications.   Whether the application of interest is the progression of an epidemic, the growth of a biological population, or the spread of a contamination plume along a river, a standard question that confronts all applied mathematicians is how to best choose model parameters to calibrate a particular mathematical model to a set of incomplete, sparse data.     

In many practical scenarios we are interested in generating both point estimates of model parameters, as well as quantifying the uncertainty in those point estimates. Quantifying uncertainty in parameter estimates is important so that we can understand how data availability and data variability impacts our ability to precisely estimate model parameters.  Understanding the extent to which parameter estimates are constrained by the quality and quantity of available data relates to the concept of \textit{parameter identifiability} which, as we will demonstrate, is a key concept that is often overlooked~\cite{Hines2014}.  While concepts of parameter estimation and parameter identifiability are dealt with in the applied statistics literature~\cite{Bates1988,Cole2020,Wasserman2004}, the kinds of mathematical models often used to demonstrate these ideas within the applied statistics literature (e.g. nonlinear regression models) may often seem unrelated to the kinds of mathematical models that are used by applied mathematicians.  In particular, we note that a broad range of practical problems are often modeled using mathematical models that are based ordinary differential equations (ODEs), including initial value problems (IVPs) and boundary value problems (BVPs), as well as partial differential equations (PDEs).

Identifiability is a property which a model must satisfy for precise parameter inference, and identifiability analysis refers to a group of methods used to determine how well the parameters of a mathematical model can be estimated given the observed data~\cite{Cobelli1980}, which may vary widely in quality and quantity across different practical applications. Methods of identifiability analysis are typically classified in terms of whether they deal with \textit{structural} or \textit{practical} identifiability~\cite{Hines2014}. Structural identifiability focuses on the question of whether different parameter values generate different probability distributions of the observable variables~\cite{Meshkat2009,Meshkat2015,Simpson2024b}. The implication of this is that, with access to an infinite amount of ideal, noise-free data, it would then be possible to precisely estimate the model parameters. As such, structural identifiability is solely concerned with analyzing the structure of the mathematical model.  Structural identifiability is often assessed using software that typically use Lie derivatives to generate a system of input–output equations, and the solvability properties of this system provide information about structural identifiability~\cite{Chis2011,Díaz-Seoane2022,Ligon2018}. In contrast, practical identifiability analysis involves working with noisy, incomplete data, and exploring the extent to which parameter values can be confidently estimated, given these data. In particular, it usually entails fitting a mathematical model to data and then exploring the extent to which the fit of the model to the data changes as the parameters are varied. Such practical identifiability analysis can be performed either locally near a given point, such as near the parameter values that provide the best model fit, or globally over the extended parameter space.  A common tool for assessing practical identifiability is the profile likelihood~\cite{Frohlich2014,Raue2009,Simpson2024b}, which is the main focus of this article.

This article aims to bridge the gap between practical mathematical modeling and parameter identifiability, parameter inference and model prediction through a series of informative computational exercises.  These exercises aim to illustrate a range of simple methods to explore parameter identifiability, parameter estimation and model prediction from the point of view of an applied mathematician.  In particular, we develop likelihood-based methods and illustrate how these flexible methods can be adapted to deal with a range of mathematical modeling frameworks such as working with ODE-based models, including both IVPs and BVPs, as well as working with PDE-based models.  We provide open source code to replicate all exercises, and we encourage readers to use this code directly or to adapt it as required for different types of mathematical models.

This article is arranged in the format of three self-contained computational exercises that relate to three different classes of mathematical models.  Section~\ref{sec:ODE} explores identifiability, estimation and  prediction for a very familiar linear ODE model where many concepts are developed visually, before they are further developed in a more general computational framework. Section`\ref{sec:PDE} explores related concepts for a PDE model, where we illustrate the consequences of using different \textit{noise models} together with the same PDE model to describe the underlying process of interest.  Results in Section~\ref{sec:ODE} and Section~\ref{sec:PDE} deal with identifiable problems whereas in Section~\ref{sec:BVP} introduces a seemingly simple BVP where our computational tools indicate that the parameters are not identifiable.  In this case we explore how a simple re-parameterization  of the likelihood function allows us to re-cast the problem in terms of identifiable parameter combinations.  Finally, in Section~\ref{sec:conclusion} we discuss options for extensions.

\section{Modeling with ODEs} \label{sec:ODE}
To develop and demonstrate key ideas we first consider a very simple mathematical model that describes the cooling (or heating) of some object at uniform temperature $T(t)$, where the uniform temperature can vary with time, $t$.  The object, initially at temperature $T(0)$, is placed into an environment of constant ambient temperature, $T_{\rm{a}}$.  Heat conduction leads to $T(t)$ increasing if $T_{\rm{a}} > T(0)$ or decreasing if $T_{\rm{a}} < T(0)$.  This heat transfer process is often modeled using Newton's law of cooling
\begin{equation}
\dfrac{\textrm{d}T(t)}{\textrm{d}t}=- k \left(T(t) - T_{\rm{a}}\right), \, \textrm{with solution } \,  T(t) = \left(T(0) - T_{\rm{a}} \right)\textrm{exp}\left(-kt\right) + T_{\rm{a}}, \label{eq:NewtonODE}
\end{equation}
where $k > 0$ is a constant heat transfer coefficient that depends upon the material properties of the object.  A classical textbook application of this model is to describe the cooling of an object (e.g.~a loaf of bread) that is removed from an oven at temperature $T(0)$, and placed into a room with ambient temperature $T_{\rm{a}}$, where $T(0) > T_{\rm{a}}$.  To use this model to describe the cooling process we must know the initial temperature $T(0)$, which for simplicity we will take to be a known constant given by the oven temperature.  We also need to know the ambient temperature $T_{\rm{a}}$ and the heat transfer coefficient $k$.  Taken together, this means that we have two unknown parameters $\vec{\theta} = (T_{\rm{a}}, k)^\mathsf{T}$ that we wish to estimate from experimental measurements.

\begin{figure}[htp]
  \centering
\includegraphics[width=1.0\textwidth]{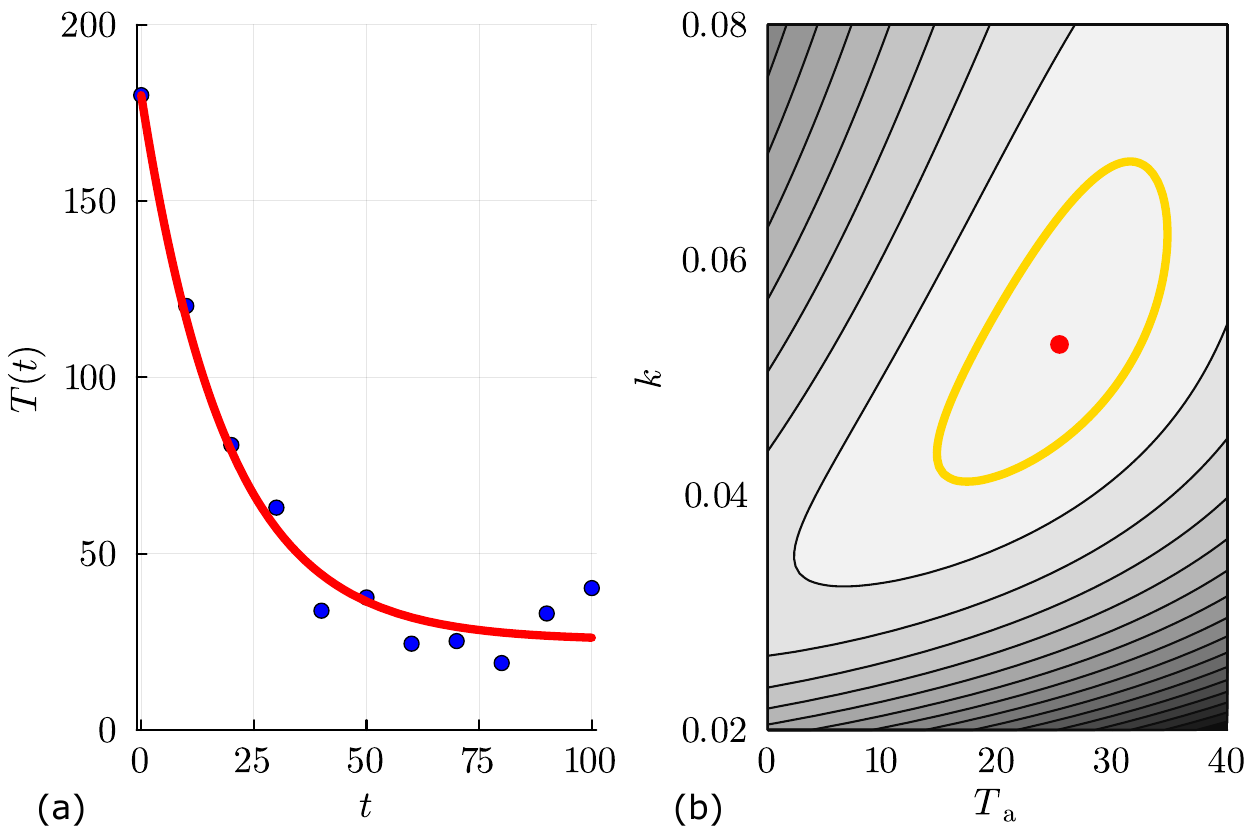}
  \caption{(a) Synthetic data (blue dots) showing observations $T^{\rm{obs}}(t)$ at $t = 0, 10, 20, \ldots, 100$ superimposed with the solution (solid red) evaluated at the maximum likelihood estimate (MLE) $\hat{\vec{\theta}} = (\hat{T_{\rm{a}}},\hat{k})^\mathsf{T} = (25.386,0.053)^\mathsf{T}$. (b)  Heat map of $\bar{\ell}(\vec{\theta} \mid T^{\rm{obs}})$ superimposed with the MLE (red dot) and a contour at $\bar{\ell}^{*} = -\Delta_{0.95,2}/2 =  -2.996$ (solid gold).  The greyscale shading darkens with decreasing $\bar{\ell}$.  Data in (a) are  obtained by solving Equation~\eqref{eq:NewtonODE} with $\vec{\theta} =  (20,0.05)^\mathsf{T}$ and corrupting the solution at  $t = 0, 10, 20, \ldots, 100$ with additive Gaussian noise with $\sigma =8$.  The units of temperature are $^{\circ}$C; time is measured in minutes; and the dimensions of $k$ are /minute. \label{fig:F1}}
\end{figure}

Figure \ref{fig:F1}(a) shows some synthetic data describing the cooling of an object from $T(0) = 180^{\circ}$C over a period of 100 minutes, where noisy measurements are made at $t=0, 10, 20, \ldots, 100$ minutes.  Later in this section we will explain how these synthetic data were generated, but for the moment it is important to note that these data are incomplete since we have just 11 discrete measurements over a 100-minute time interval, and our visual interpretation of the data indicates they are noisy in the sense that we see clear fluctuations in the measurements about the overall decreasing trend.  We will attribute these fluctuations to some kind of measurement error in the data generation process.

This incomplete, noisy data motivates us to ask four natural questions:
\begin{enumerate}
    \item What value of $\vec{\theta} = (T_{\rm{a}}, k)^\mathsf{T}$ in  Equation~\eqref{eq:NewtonODE} gives $T(t)$ that provides the best match to the data? 
    \item How confident are we in these best-fit estimates of $\vec{\theta}$? In other words, to what extent does the incomplete data constrain our estimate of $\vec{\theta}$?
    \item How can we measure uncertainty in our estimate of $\vec{\theta}$?
    \item How does the uncertainty in our estimate of $\vec{\theta}$ propagate into uncertainty in our ability to predict $T(t)$ in a future experiment?  
\end{enumerate}

\medskip\noindent
The aim of this article is to outline a simple, yet powerful approach to address these questions using standard mathematical and computational tools.  Some relatively basic background knowledge, typically covered in the first two years of most undergraduate mathematics curricula, is assumed.   For example, one way of interpreting and understanding what we mean by the best-fit estimate of $\vec{\theta}$ (above) is to think of a standard least-squares procedure for finding the best-fitting model solution to a given set of data by minimizing the sum of the residuals, defined to be the squares of the offsets of the data from the model solution.  We will elaborate more on this point below.  Another key ingredient of our work is to deal with a range of commonly-encountered univariate probability distributions.  These distributions, such as the normal and log-normal distributions, are familiar functions of one variable that give the probabilities of occurrence of possible outcomes for an experiment.  Perhaps the most advanced mathematical concept we rely on is numerical optimization~\cite{Audet,Nocedal}.  For all calculations in this work we use the NLopt library~\cite{Johnson2024} where the Nelder-Mead algorithm is implemented with simple bound constraints.  We have chosen to work with the Nelder-Mead algorithm within NLopt because this is a standard option for numerical optimization that is very well tested and understood~\cite{Audet,Nocedal}, however we acknowledge that it is also be possible to work with a different algorithm to implement the numerical optimization steps.

In this context we will refer to Newton's law of cooling, as stated mathematically in Equation~\eqref{eq:NewtonODE}, as a \textit{process model} because this mathematical model describes the process of interest (i.e. heat transfer).  To proceed we also introduce a \textit{noise} model which relates the observed data $T^{\rm{obs}}(t)$ to the solution $T(t)$ of the process model.   In this first example we make a standard assumption that the observed data, $T^{\rm{obs}}(t)$, are samples from a normal distribution where the mean of that distribution is the solution of the process model.  This is commonly referred to as additive Gaussian noise in the literature. Under this standard approach, at any time $t$ we have $T^{\rm{obs}}(t) \mid \vec{\theta} \sim \mathcal{N}\left(T(t), \sigma^2\right)$, where $T(t)$ is temperature at time $t$, as given in Equation~\eqref{eq:NewtonODE}, and $\sigma^2$ is a constant variance. Throughout this work we treat the variance as a fixed constant, and we will comment on our choice of $\sigma^2$ later. This noise model provides a means of relating the solution of the process model to the observed data through the probability density function of the normal distribution.  For example, suppose we have measured some value of $T^{\rm{obs}}(\tau)$ at time $t = \tau$.  Within this framework we can compute the density of various predictions using properties of the normal distribution since we have $T^{\rm{obs}}(\tau)~\mid~\vec{\theta}~\sim~\mathcal{N}\left(T(\tau), \sigma^2\right)$.  In the simple case of having a single measurement it is clear that the value of $T(\tau)$ that best matches the single measurement is  $T(\tau) = T^{\rm{obs}}(\tau)$.  If $T(\tau) \ne T^{\rm{obs}}(\tau)$, we can quantify this in a probabilistic sense in terms of the probability density function. For example, consider a scenario where we have $T^{\rm{obs}}(\tau) = 10$, with a fixed value of $\sigma^2 = 1$. If a particular choice of $\vec{\theta}$ leads to $T(\tau) = 10$, the density is $\phi(10;10,1) = 0.399$, where $\phi(x; \mu, \sigma^2)$ denotes the probability density of observing $x$ given the normal distribution with mean $\mu$ and variance $\sigma^2$. Alternatively, if a different choice of $\vec{\theta}$ leads to $T(\tau) = 8$, the density is $\phi(8;10,1) = 0.054$, which provides a probabilistic measure of observing the data given the parameters.

For a series of measurements at time $t_i$ for $i=1,2,3,\ldots, I$, it is unreasonable to expect that the solution of the process model will perfectly match all observations simultaneously. One way of interpreting the noise model is that it represents independent fluctuations in the data gathering process.  Therefore, invoking an independence assumption means that we can evaluate the probability density at each of the $I$ measurements since $T^{\rm{obs}}(t_i) \mid \vec{\theta} \sim \mathcal{N}\left(T(t_i), \sigma^2\right)$, and taking the product of these probability densities gives us a quantity, called the \textit{likelihood}, which measures how well the model explains observed fixed data by calculating the probability of seeing that fixed data under different parameter values of the model. As we might anticipate, taking a product like this can lead to extremely small numerical quantities which can be circumvented by taking logarithms, giving rise to the log-likelihood which can be written as 
\begin{equation}\label{eq:GaussianLikelihood}
	\ell(\vec{\theta} \mid T^{\rm{obs}}(t)) = \sum_{i=1}^{I} \log \left[\phi\left(T^{\rm{obs}}(t_i); T(t_i), \sigma^2 \right)\right],
\end{equation}
where, as before, $\phi(x; \mu, \sigma^2)$ denotes the probability density function of the normal distribution with mean $\mu$, variance $\sigma^2$, and $T(t_i)$ is the temperature at time $t = t_i$ for $i=1,2,3,\ldots, I$, as given in Equation~\eqref{eq:NewtonODE}. In the case of working with an additive Gaussian noise model with constant variance, we can re-write the right hand side of Equation~\eqref{eq:GaussianLikelihood} in terms of the normal probability density function to give
\begin{equation}\label{eq:GaussianLikelihood2}
	\ell(\vec{\theta} \mid T^{\rm{obs}}(t)) =   -\dfrac{I}{2}\log \left(2\pi \sigma^2\right) - \dfrac{1}{2\sigma^2}\sum_{i=1}^{I} \left(T^{\rm{obs}}(t_i) - T(t_i) \right)^2.
\end{equation}
Recalling that $\sigma^2$ is a constant, writing the log-likelihood function in this way 
confirms that maximizing $\ell$ is mathematically equivalent to minimizing the standard least-squares objective function~\cite{Hines2014}, noting that the first term on the right of Equation~\eqref{eq:GaussianLikelihood2} is a constant.

Given the log-likelihood function we may now address the first question (above) using numerical optimization to estimate the value of $\vec{\theta}$, denoted $\hat{\vec{\theta}}$, that maximizes the log-likelihood, $\sup_{\vec{\theta}} \ell(\vec{\theta} \mid T^{\rm{obs}}(t))$. For simple problems  with one or two unknown parameters we can visualize this maximization simply by plotting $\ell(\vec{\theta} \mid T^{\rm{obs}}(t))$ as a function of $\vec{\theta}$ and visually identifying the value of $\vec{\theta}$ that maximizes $\ell(\vec{\theta} \mid T^{\rm{obs}}(t))$.  For more complicated problems, with three or more unknown values, this graphical approach is infeasible so we use numerical optimization.  Figure~\ref{fig:F1}(b) shows a filled greyscale contour plot of $\ell((T_{\textrm{a}},k)^\mathsf{T} \mid T^{\rm{obs}}(t))$ where we see that a single value of $\vec{\theta}$  maximizes the log-likelihood function.  In this case numerical optimization gives $\hat{\vec{\theta}} = (25.386,0.053)^\mathsf{T}$  which is the  \textit{maximum likelihood estimate} (MLE).  Evaluating~\eqref{eq:NewtonODE} at the MLE, and superimposing the MLE solution onto the data in Figure~\ref{fig:F1}(a) indicates that this solution provides a good visual match to the data. 

Our point estimate of $\hat{\vec{\theta}}$ gives us the value of $\vec{\theta}$ that means that $T(t)$ is the best match to the data, in terms of minimizing the sum-squared error, but this point estimate does not provide any indication of the uncertainty in our estimate of $\vec{\theta}$. For example, to what extent can other parameter sets provide almost as good a fit to the data?  To address our second question (above) we will work with the normalized log-likelihood function 
\begin{equation}\label{eq:GaussianLikelihoodN}
	\bar{\ell}(\vec{\theta} \mid T^{\rm{obs}}(t)) = \ell(\vec{\theta} \mid T^{\rm{obs}}(t)) - \ell(\hat{\vec{\theta}} \mid T^{\rm{obs}}(t)),
\end{equation}
so that we have $\bar{\ell}(\hat{\vec{\theta}} \mid T^{\rm{obs}}(t)) =0$.

The key to inferential precision is the curvature of the log-likelihood function.  Intuitively we expect that if $\bar{\ell}(\vec{\theta} \mid T^{\rm{obs}}(t))$ is tightly peaked near $\hat{\vec{\theta}}$ then the data constrains our parameter estimates to a relatively narrow region in parameter space (since small changes in the parameter values entail large changes in the log-likelihood).  In contrast, if $\bar{\ell}(\vec{\theta} \mid T^{\rm{obs}}(t))$ is relatively flat near $\hat{\vec{\theta}}$ then the data contains insufficient information to constrain our estimate of $\vec{\theta}$ and it is possible that the solution of the model with many values of $\vec{\theta}$ can accurately match the data.  

The degree of curvature of the log-likelihood function can be graphically assessed for mathematical models involving just one or two parameters, but more generally we use numerical optimization together with the concept of the \textit{profile likelihood}~\cite{Pawitan2001} to provide insight where simple visualization is not possible.

\begin{figure}[htp]
  \centering  \includegraphics[width=1.0\textwidth]{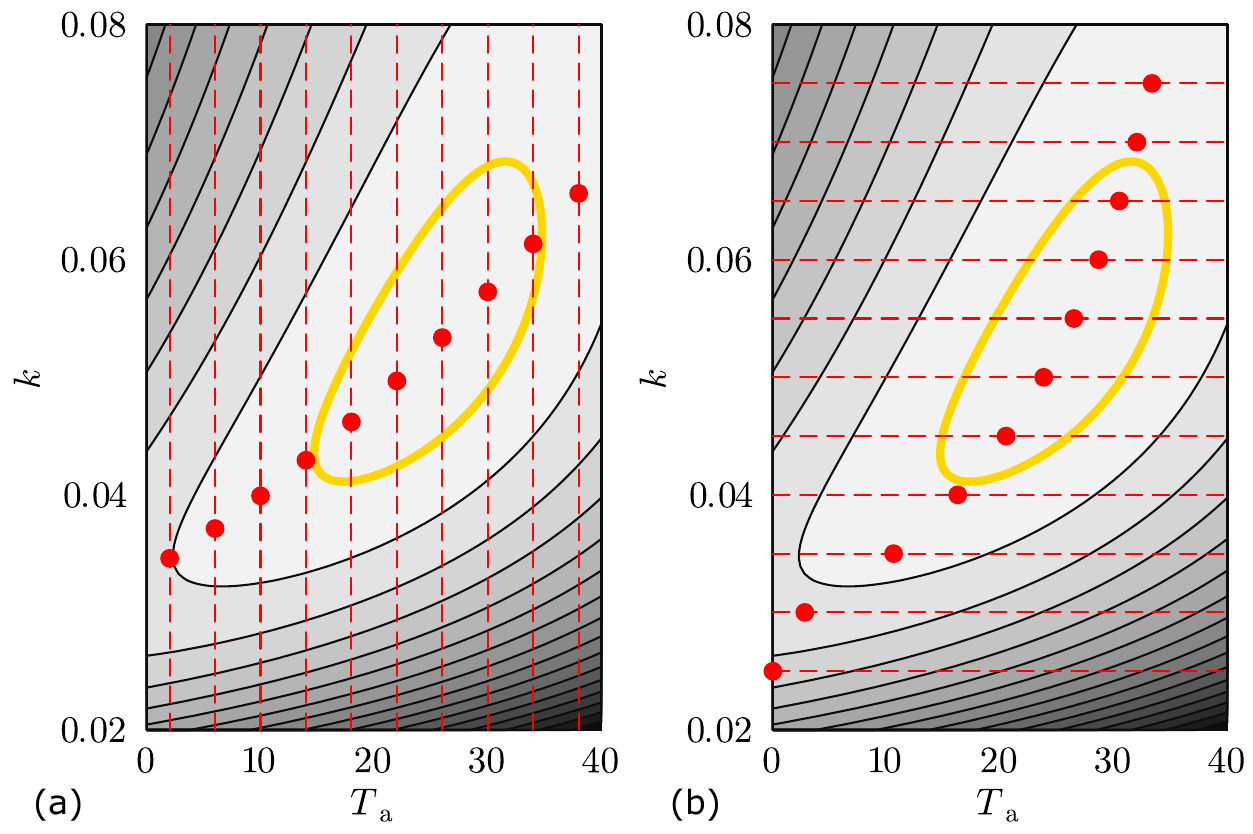}
  \caption{Visualization of the profile likelihood calculation for: (a) $T_{\rm{a}}$  and (b) $k$.  Each subfigure shows the same heat map of $\bar{\ell}$ and 95\% threshold shown previously in Figure \ref{fig:F1}(b). In (a) we consider 10 equally-spaced values of the interest parameter $T_{\rm{a}} = 2, 4, \ldots, 38$, as indicated by the vertical dashed lines.  Along each vertical line we identify the value of $k$ for which $\bar{\ell}$ is maximized (red dots).  Similarly, in (b) we consider 11 equally-spaced values of the interest parameter $k = 0.025, 0.030, \ldots, 0.075$, as indicated by the horizontal dashed lines.  Along each horizontal line we identify the value of $T_{\rm{a}}$ for which $\bar{\ell}$ is maximized (red dots). \label{fig:F2}}
\end{figure}

\begin{figure}[htp]
  \centering  \includegraphics[width=1.0\textwidth]{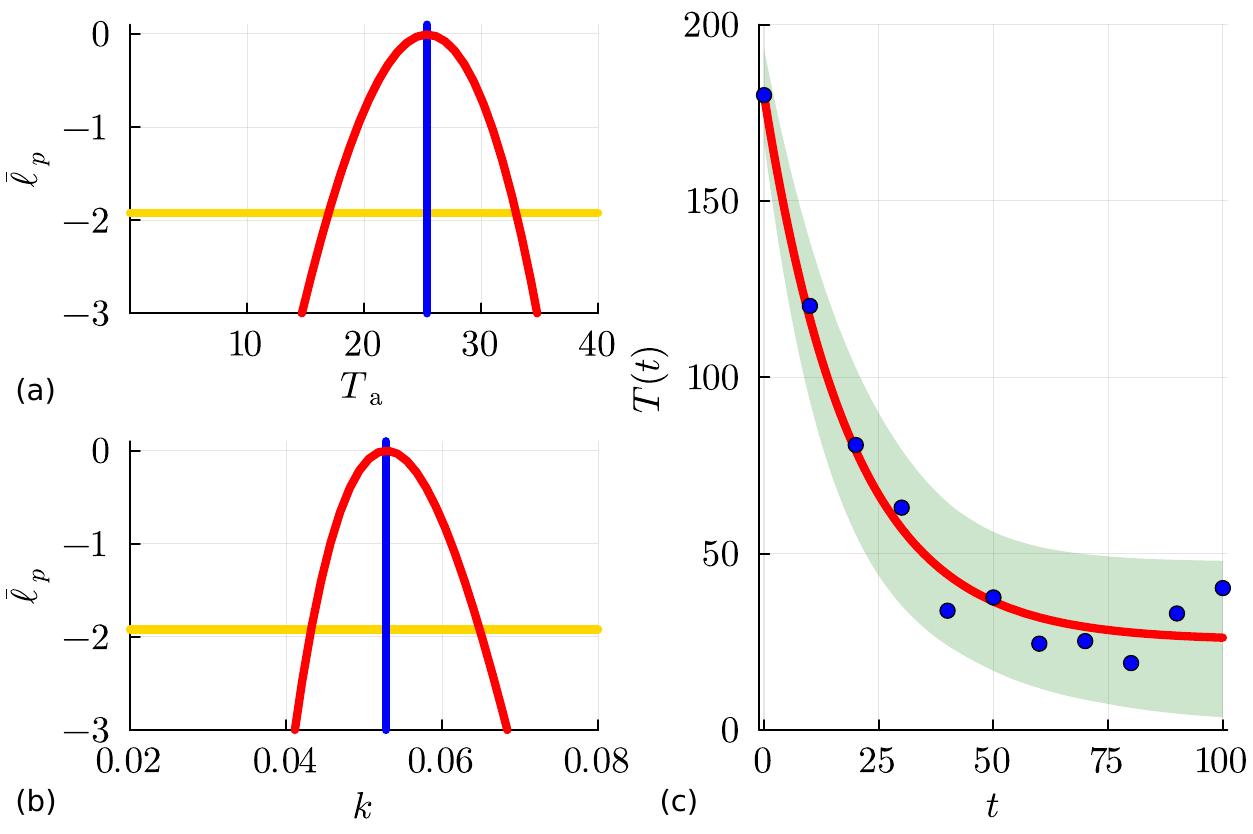}
  \caption{(a)--(b) Univariate profile likelihood functions for $T_{\rm{a}}$ and $k$, as indicated (solid red).  Each profile indicates the MLE (solid blue) and the $95\%$ threshold $\bar{\ell}^{*} = -\Delta_{0.95,1}/2  =  -1.921$ contour (solid gold).  The 95\% confidence intervals are $T_{\rm{a}}\in[16.977,32.983]$ and $k\in[0.0432,0.0648]$.  Profile likelihood function for $T_{\rm{a}}$ is computed using a uniform mesh $T_{\rm{a}} = 1,2,3,\ldots, 40$, and the profile likelihood function for $k$ is computed using a uniform mesh $k = 0.02, 0.04, \ldots, 0.10$. (c) Comparison of data (blue dots), MLE solution (solid red) and a 95\% prediction interval (green shaded region). The units of temperature are $^{\circ}$C; time is measured in minutes; and the dimensions of $k$ are /minute. \label{fig:F3}}
\end{figure}

Profile likelihood functions have a straightforward interpretation that can be explained in terms of the two-dimensional contour plots of $\bar{\ell}$ in Figure \ref{fig:F1}(b). We will now explain how we compute profile likelihood functions by making reference to this contour plot.  First, we treat $T_{\rm{a}}$ as an \textit{interest parameter} and consider a relatively coarse uniform discretisation of that parameter with 10 equally-spaced mesh points at $T_{\rm{a}} = 2, 6, \ldots, 38$.  We take a relatively coarse discretisation of $T_{\rm{a}}$ for ease of illustration and visualization, and later we will refine our calculations using a finer mesh. For each value of $T_{\rm{a}}$ we draw a vertical line across the contour plot of $\bar{\ell}$ and identify the value of $k$ where $\bar{\ell}$ is maximized along that vertical straight line.  For example, in Figure \ref{fig:F2}(a) along the vertical line where $T_{\rm{a}} = 10$, the maximum value of $\bar{\ell}$ is $-5.780$, and this maximum occurs at $k =0.040$.  Repeating this optimization for each value of $T_{\rm{a}}$ over the grid of $T_{\rm{a}}$ values reduces the two-dimensional log-likelihood function to a univariate function, called the profile likelihood~\cite{Pawitan2001}, which we denote $\bar{\ell}_p$.  This univariate profile likelihood function can be used qualitatively and quantitatively to assess the curvature of the log-likelihood function at the MLE.  Repeating this process of holding the interest parameter constant across a finer uniform mesh and optimizing out the nuisance parameter allows us to construct the profile likelihood function for $T_{\rm{a}}$ shown in Figure \ref{fig:F3}(a) where we simply plot $\bar{\ell}_p$ as a function of the interest parameter,  $T_{\rm{a}}$.

The process of holding one interest parameter constant and optimizing out the remaining nuisance parameter(s) by maximizing $\bar{\ell}$ can be repeated for each component of $\vec{\theta}$.  In the current problem we have two parameters and so we can construct two univariate profile likelihood functions.  We now treat $k$ as the interest parameter by considering a relatively coarse uniform discretisation with 11 equally-spaced mesh points, $k = 0.025, 0.030, \ldots, 0.075$.  For each value of $k$ we draw a horizontal line across the contour plot of $\bar{\ell}$ in Figure \ref{fig:F2}(b) and identify the value of $T_{\rm{a}}$ where $\bar{\ell}$ is a maximum along that horizontal straight line.  For example, in Figure \ref{fig:F2}(b) for $k = 0.05$, the maximum value of $\bar{\ell}$ is $-0.141$, and this occurs at $T_a = 23.832$.  Repeating this optimization for each value of $k$ over a uniform grid reduces the two-dimensional log-likelihood function to a univariate function. Repeating this process of holding the interest parameter constant across a finer uniform mesh and optimizing out the nuisance parameter allows us to construct the profile likelihood function for $k$ shown in Figure \ref{fig:F3}(b). 

For our ODE problem with two unknown parameters the process of fixing one interest parameter and optimizing out the other nuisance parameter has a simple graphical interpretation that we have made explicit in Figure \ref{fig:F2}.  This visual interpretation is unclear when we consider dealing with higher dimensional problems with three or more parameters.  In general we use the same numerical optimization approach to construct univariate profile likelihood functions since this approach holds regardless of the number of unknown parameters.  Both univariate profile likelihood functions in Figure \ref{fig:F3} involve a single, well-defined peak at the MLE indicating that these parameters are identifiable.  

The degree of curvature of the log-likelihood function can be quantified in several ways.  In this work we take a very simple approach by identifying a threshold-based interval for each parameter defined by the interval where the $\bar{\ell}_p \ge \bar{\ell}^{*}$, where the threshold log-likelihood value is  associated with an asymptotic confidence interval~\cite{Royston2007}.   The threshold profile log-likelihood value can be approximately calibrated using the $\chi^2$ distribution, leading to $\bar{\ell}^{*} = -\Delta_{n,q}/2$ where $\Delta_{n,q}$ refers to the $q$th quantile of a $\chi^2$ distribution with $n$ degrees of freedom, taken to be the dimension of the parameter of interest (i.e. the number of interest parameters)~\cite{Royston2007}.  For example, with the univariate profile likelihood functions (where the dimension of the interest parameter is one) we can identify a 95\% confidence interval with the threshold of $\bar{\ell}^{*} = -\Delta_{1,0.95}/2 = -1.912$.  For $T_{\rm{a}}$ we have $\hat{T}_{\rm{a}} = 25.386$, and the 95\% confidence interval is $T_{\rm{a}}\in[16.977, 32.983]$; for $k$ we have $\hat{k} = 0.053$, and the 95\% confidence interval is $k\in[0.0432,0.0648]$. These univariate intervals provide a simple, convenient measure of the uncertainty in each parameter.  These confidence intervals are derived using the full log-likelihood function and the relevant asymptotic thresholds, and profiling allows us to obtain confidence intervals for each parameter, one-at-a-time.

Results in Figures \ref{fig:F1}-\ref{fig:F3} have answered the first three of four questions (above), confirming that we can estimate the best-fit parameters, and establish our uncertainty in these estimates. We now turn to examining how our uncertainty in $\vec{\theta}$ can be related to the predictive uncertainty in $T(t)$. The log-likelihood function in Figure \ref{fig:F1}(b) is superimposed with a contour at  threshold value $\bar{\ell}^{*} = -\Delta_{2,0.95}/2 = -2.996$.  Choices of $\vec{\theta}$ within this contour are contained within the asymptotic 95\% confidence set of $\vec{\theta}$ whereas choices outside of this contour are not.

To explore how variability of $\vec{\theta}$ within this confidence set translates into variability in predictions of $T(t)$ we can randomly sample values of $\vec{\theta} = (T_{\rm{a}},k)^\mathsf{T}$ to generate a set of $M$ samples within the confidence set.  For each of the $M$ samples we evaluate Equation~\eqref{eq:NewtonODE} to give $M$ continuous solution curves, $T_m(t)$ for $m=1,2,3,\ldots,M$.  According to the noise model, these $M$ solutions describe the \textit{curve-wise} mean of the noise model distribution.  To provide a measure of the variability about each mean trajectory we can use properties of the noise model to quantify and visualize the variability in $T(t)$ about the mean.  A standard way to characterize the width of a probability distribution is to consider various quantiles of that distribution.  

Using these ideas we can describe the variability about the mean trajectories noting that with $\sigma = 8$ the 5\% and 95\% quantiles define a curve-wise interval $T_m(t) \pm 13.159$ for each of the $M$ trajectories, $m=1,2,3,\ldots,M$.  To provide an overall prediction interval that accounts for variability associated with the noise model and the variability introduced by considering different choices of $\vec{\theta}$ within the confidence set we evaluate $T_m(t)$ at $t=0,1,2,3,\ldots,100$ minutes for each of the $M$ trajectories.  We than record the maximum and minimum values of of $T_m(t) \pm 13.159$ across all $M$ trajectories evaluated at  $t=0,1,2,3,\ldots,100$ minutes to give the prediction interval in Figure \ref{fig:F3}(c).  This prediction interval provides a quantitative indication of how variability in $\vec{\theta}$ maps to variability in $T(t)$~\cite{Simpson2024a}.  The prediction interval in Figure \ref{fig:F3} is obtained using $M=1000$ random samples.  In this case this choice of $M$ is sufficiently large that the prediction interval is visually insensitive to taking more samples of $\vec{\theta}$.

Throughout this exercise we have treated $\sigma^2$ as a constant, and our choice for this constant was made for pedagogical reasons so that the variability in the data in Figure \ref{fig:F1}(a) is visually obvious.  We then explored how this variability in the data propagated into the shape of the log-likelihood function, the width of the profile likelihood functions, and ultimately the width of the prediction intervals.  We encourage readers to explore repeating these calculations by generating different sets of data with different $\sigma^2$ to investigate how higher-quality data (i.e. smaller $\sigma^2$) leads to narrower profile likelihood functions and prediction intervals, as well as investigating the consequence of working with lower-quality data (i.e. larger $\sigma^2$).  These explorations are  straightforward using the software provided.

\section{Modeling with PDEs} \label{sec:PDE}
\begin{figure}[htp]
  \centering
  \includegraphics[width=0.8\textwidth]{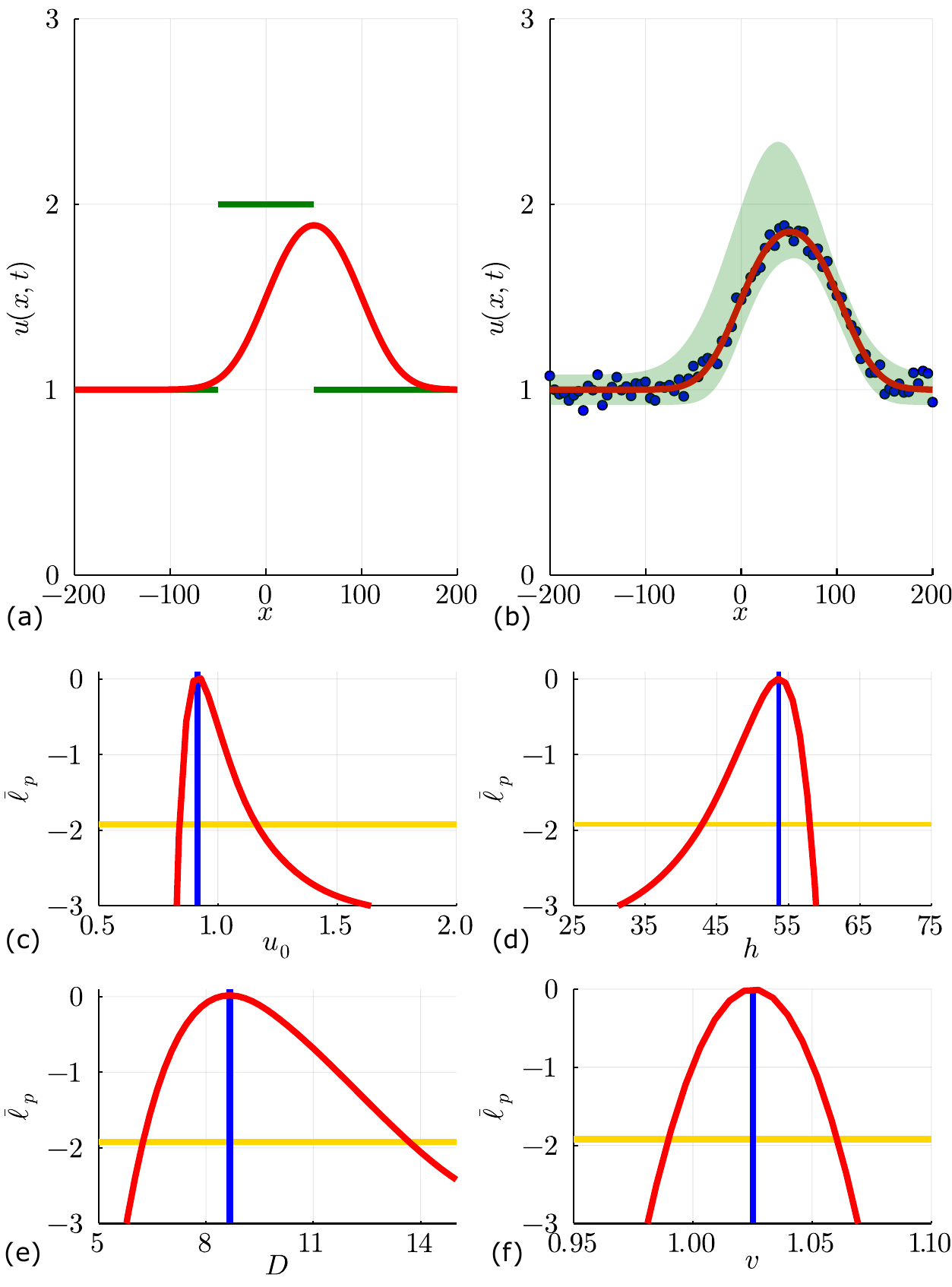}
  \caption{(a) Initial condition $u(x,0)$ (solid green) with $u_{\textrm{b}}=u_0=1$ and $h=50$ superimposed with the PDE solution at $t=50$, $u(x,50)$ (solid red) with $D=10$ and $v=1$.  (b) Synthetic data (blue dots) showing observations $u^{\rm{obs}}(x,50)$ at $x = -200, -195, -190, \ldots, 195, 200$ superimposed with the MLE solution (solid red) with $\hat{\vec{\theta}} = (\hat{u_0},\hat{h},\hat{D},\hat{v})^\mathsf{T} = (0.914,53.693,8.671,1.025)^\mathsf{T}$ and the 95\% prediction interval (green shaded region). The data are obtained by evaluating Equation~\eqref{eq:ADEPDE1} with $(u_0,h,D,v)^\mathsf{T} = (1, 50, 10, 1)^\mathsf{T}$ at $x = -200, -195, -190, \ldots, 195,  200$ and corrupting each data point with additive Gaussian noise with $\sigma = 0.05$. (c)--(f) Profile likelihood functions for $u_0$, $h$, $D$ and $v$, as indicated (solid red).  Each profile indicates the MLE (solid blue) and the $95\%$ threshold $\bar\ell^{*} = -\Delta_{0.95,1}/2 =  -1.921$  (solid gold). The 95\% confidence intervals are $u_0\in[0.839,1.156]$, $h\in[42.939,58.043]$, $D\in[6.227,13.686]$ and $v\in[0.990,1.060]$. \label{fig:F4}}
\end{figure}

With this example we will demonstrate how the concepts developed in Section \ref{sec:ODE} can be adapted to apply to mathematical models based on PDEs by working with the advection-diffusion equation to describe the spatio-temporal distribution of a non-dimensional concentration $u(x,t) \ge 0$,
\begin{equation}
\dfrac{\partial u}{\partial t} =  D \dfrac{\partial^2 u}{\partial x^2} - v \dfrac{\partial u}{\partial x}, \quad \textrm{on} \quad  -\infty < x < \infty, \label{eq:ADEPDE1}
\end{equation}
where $D > 0$ [L$^2$/T] is the diffusion coefficient, and $v$ [L/T] is the advective velocity.  We have not specified the units of $D$ or $v$, instead we use standard notation~\cite{Barenblatt2003} to indicate that the diffusion coefficient has dimensions of length squared per time, and the advective velocity has dimensions of length per time. This mathematical model is widely used in various physical, chemical and biological applications, such as the study of the dispersion of dissolved solutes (e.g.~nutrients, pollutants) in porous media~\cite{Zheng2002}.   We will consider the solution of Equation~\eqref{eq:ADEPDE1} for the initial condition $u(x,0) = u_{\textrm{b}}+u_0$ for $|x| < h$, and $u(x,0) = u_{\textrm{b}}$ for $|x| > h$, where $h > 0$ is the half-width of the initial condition about the origin. We interpret $u_{\textrm{b}} \ge 0$ as a uniform \textit{background} concentration of $u$, and we are particularly interested in the spatial spreading of $u$ that arises when an additional amount of $u$ is placed uniformly within distance $h>0$ from the origin, giving $u(x,0) = u_{\textrm{b}} + u_0$ for $|x| < h$, where $u_0 > 0$. This initial condition with $h=50$ is plotted in Figure \ref{fig:F4}(a) where we have a background concentration $u_{\textrm{b}}=1$ in the far-field, and locally within the interval $-50 < x < 50$ we have $u(x,0)= 2$ (i.e.~$u_0=1$).  The solution of the mathematical model describes how the initial density profile evolves as a result of combined advection and diffusive transport as a function of position $x$ and time $t$.

The solution of Equation~\eqref{eq:ADEPDE1} with these initial conditions can be obtained using a Fourier transform and is given by 
\begin{equation}
u(x,t) =  u_{\textrm{b}} + \dfrac{u_0}{2}\left[\textrm{erf}\left(\dfrac{h-(x-vt)}{2\sqrt{Dt}}\right)+\textrm{erf}\left(\dfrac{h+(x-vt)}{2\sqrt{Dt}}\right)  \right], \label{eq:ADEPDESolution}
\end{equation}
where $\textrm{erf}(x) = \left(2/\sqrt{\pi}\right)\int_{0}^{x}\textrm{exp}(-z^2) \; \textrm{d}z$ is the error function~\cite{Abramowitz1964}.  Results in Figure \ref{fig:F4}(a) superimpose the exact solution for $(u_0,h,D,v)^\top = (1, 50, 10, 1)$ at $t=50$ onto the plot of $u(x,0)$ and we see that the center of mass of $u(x,50)$ translates in the positive $x$-direction from $x=0$ to $x=50$ as a result of the advective transport.  In addition, the discontinuous $u(x,0)$ profile becomes continuous and smooth by $t=50$ owing to the action of diffusive transport.  

Discrete data, shown in Figure \ref{fig:F4}(b), is obtained by evaluating the exact solution, $u(x,50)$ at $x=-200, -195, -190, \ldots, 200$ and then corrupting each value of $u(x,50)$ with additive Gaussian noise with $\sigma = 0.05$.  Given this noisy data we will now address the same questions of parameter estimation, parameter identifiability and model prediction as in Section \ref{sec:ODE} except now we are dealing with four parameters $\vec{\theta} = (u_0,h,D,v)^\mathsf{T}$ in the PDE model instead of dealing with just two parameters in the ODE model.  An important consequence of working with a larger number of unknown parameters is that we can no longer simply visualize the log-likelihood function as we did in Figure \ref{fig:F1}(b).

As in the ODE model, here we have a log-likelihood function
\begin{equation}\label{eq:GaussianLikelihoodADE}
	\ell(\vec{\theta} \mid u^{\rm{obs}}(x_i,t)) = \sum_{i=1}^{I} \log \left[\phi\left(u^{\rm{obs}}(x_i,t); u(x_i,t), \sigma^2 \right)\right],
\end{equation}
where again $\phi(x; \mu, \sigma^2)$ denotes the probability density function of the normal distribution with mean $\mu$ and variance $\sigma^2$.  Here, the index $i$ refers to the spatial position where measurements of density are taken. Although we are unable to visualize this function like we did in Section \ref{sec:ODE}, numerical optimization gives  $\hat{\vec{\theta}} =  (0.914,53.693,8.671,1.025)^\mathsf{T}$. Superimposing $u(x,t)$ evaluated with $\hat{\vec{\theta}}$ in Figure \ref{fig:F4}(b) indicates that the solution provides a good visual match to the data, as anticipated. Given the MLE we can now work with a normalized log-likelihood function
\begin{equation}\label{eq:GaussianLikelihoodADE}
	\bar{\ell}(\vec{\theta} \mid u^{\rm{obs}}(x,t)) = \ell(\vec{\theta} \mid u^{\rm{obs}}(x,t)) - \ell(\hat{\vec{\theta}} \mid u^{\rm{obs}}(x,t)),
\end{equation}
which can be used to construct various profile likelihood functions to explore the identifiability of the four parameters.  Since we have four unknown parameters we  construct four univariate profile likelihood functions and the approach for each is the same.  For example, if we take $u_0$ to be the interest parameter and $(h,D,v)^\mathsf{T}$ to be the nuisance parameters, the profile likelihood for $u_0$ can be written as 
\begin{equation}
\bar{\ell}_p(u_0 \mid u^{\rm{obs}}(x,t)) = \sup_{(h,D,v)^\mathsf{T}} \bar{\ell}(\vec{\theta} \mid u^{\rm{obs}}(x,t)),
\end{equation}
which can be evaluated by holding $u_0$ at some fixed value $u_0^\dagger$ and computing values of $(h,D,v)^\mathsf{T}$ that maximize $\bar{\ell}$ using numerical optimization.  Repeating this process across a grid of $u_0$ gives a univariate profile likelihood as shown in Figure \ref{fig:F4}(c) where we see that the univariate profile likelihood for $u_0$ has a single peak at the MLE, $\hat{u}_0 = 0.914$, and approximate 95\% confidence intervals are $u_0\in[0.839,1.156]$.  Repeating this process to construct univariate profile likelihood functions for $h$, $D$ and $v$ leads to the profile likelihood functions given in Figure \ref{fig:F4}(d)--(f) indicating that all parameters are practically identifiable in this case. 

We conclude this exercise by returning to the full log-likelihood function and using rejection sampling to generate $M$ parameter samples $\vec{\theta}$, where $\bar{\ell} \ge \bar{\ell}^{*}$.  Evaluating $u(x,t)$ for each of the $M$ parameter samples, computing the 5\% and 95\% quantiles of the noise model at each value of $x$, and then evaluating the maximum and minimum at $x=-200, -195, -190, \ldots, 200$ over all $M$ samples gives the prediction interval in Figure \ref{fig:F4}(b) which illustrates how parameter estimates within the 95\% confidence set translate into a prediction interval in $u(x,50)$ for this problem.

This PDE example highlights an important shortcoming of working with the additive Gaussian noise model. A mathematical property of the exact solution, Equation~\eqref{eq:ADEPDESolution}, is that $u_{\textrm{b}} < u(x,t) < u_0 + u_{\textrm{b}}$ for $t > 0$ and $-\infty < x < \infty$ since the parabolic PDE model obeys a maximum principal~\cite{Protter1967}.  Our data in Figure \ref{fig:F4}(b) clearly violates this property as we have $u^{\rm{obs}}(x,t) < u_{\textrm{b}}$ at several locations.  This issue is even more concerning if we consider the realistic case of having no background concentration by setting $u_{\textrm{b}}=0$.  Under these circumstances  working with additive Gaussian noise is clearly unsatisfactory since this leads to $u^{\rm{obs}}(x,t) < 0$ which is physically impossible when $u$ represents a concentration that is non-negative by definition.  

\begin{figure}[htp]
  \centering
  \includegraphics[width=0.8\textwidth]{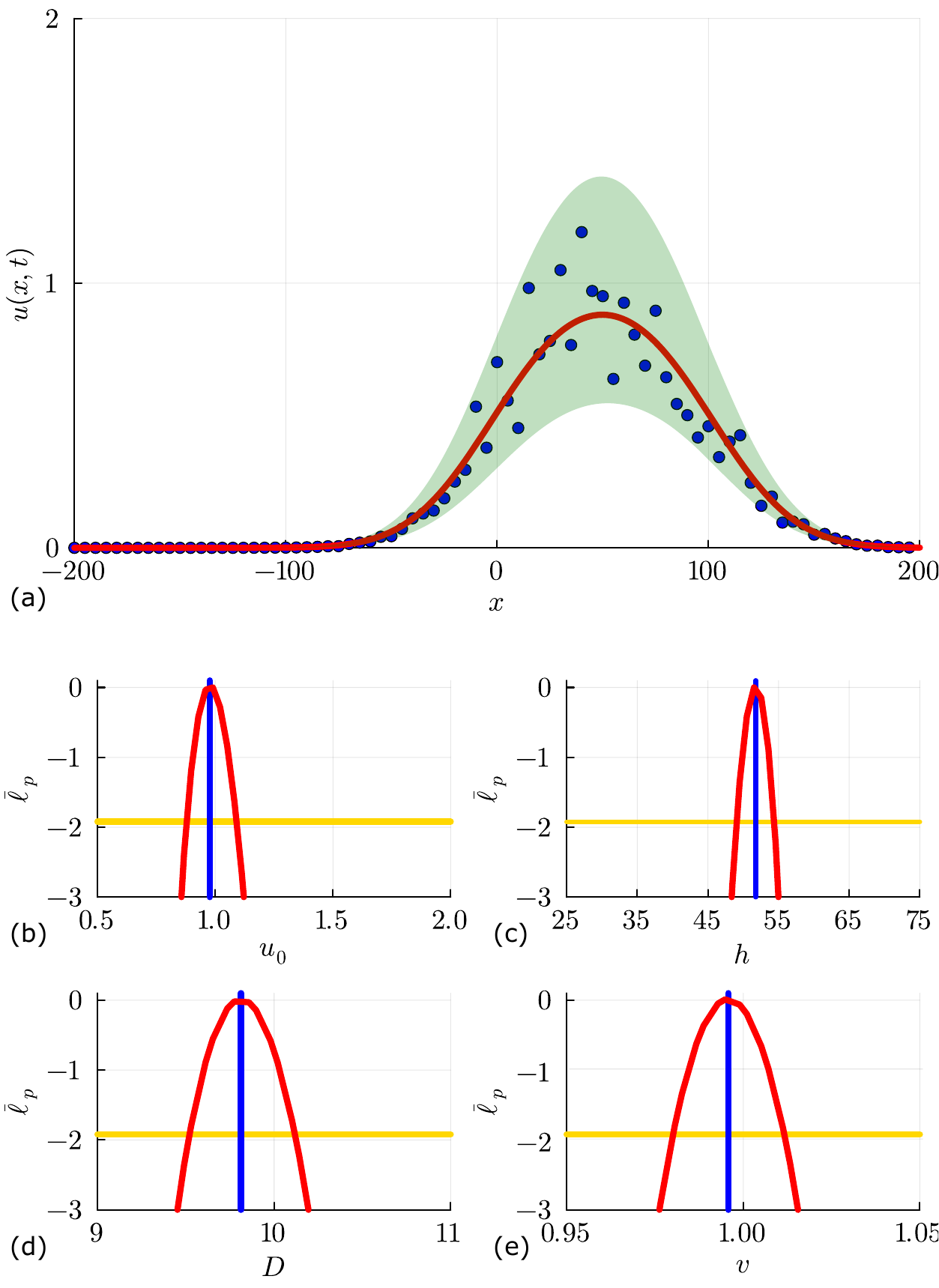}
  \caption{(a) Synthetic data (blue dots) that represent observations $u^{\rm{obs}}(x,50)$ at $x = -200, -195, -190, \ldots, 195, 200$ superimposed on the MLE solution (solid red) with $\hat{\vec{\theta}} = (\hat{u_0},\hat{h},\hat{D},\hat{v})^\mathsf{T} = (0.977,51.776,9.813,0.996)^\mathsf{T}$ and a 95\% prediction interval (green shaded region). The data are obtained by evaluating Equation~\eqref{eq:ADEPDESolution} with $(u_0,h,D,v)^\mathsf{T} = (1, 50, 10, 1)^\mathsf{T}$ at $x = -200, -195, -190, \ldots, 195, 200$ and corrupting each value of $u$ with with multiplicative log-normal noise with $\sigma = 0.2$. (b)--(e) Profile likelihood functions for $u_0$, $h$, $D$ and $v$, as indicated (solid red).  Each profile indicates the MLE (solid blue) and the $95\%$ threshold $\bar\ell^{*} = -\Delta_{0.95,1}/2 =  -1.921$  (solid gold). The 95\% confidence intervals are $u_0\in[0.879,1.090]$, $h\in[49.064,54.380]$, $D\in[9.522,10.117]$ and $v\in[0.980,1.012]$. \label{fig:F5}}
\end{figure}

One way to address this shortcoming is to work with a different noise model.  In this case we could instead introduce a multiplicative noise model where the fluctuations are no longer constant or symmetric about the mean.  For example, we could specify a noise model where the magnitude of the fluctuations is proportional to the mean of the distribution, $u^{\rm{obs}}(x_i,t) \mid \vec{\theta} = (u(x_i,t) \mid \vec{\theta})\eta_i$ where $\eta_i~\sim~\textrm{log-normal}(0,\sigma^2)$~\cite{Murphy2024}.  Our previous examples with additive Gaussian noise models have a constant variability whereas the multiplicative log-normal noise model has variability that increases with $u^{\rm{obs}}(x,t)$.  The log-normal noise model also has the attractive property that the variability vanishes as $u^{\rm{obs}}(x,t)~\to~0^+$.  Figure \ref{fig:F5}(a) shows a solution of Equation~\eqref{eq:ADEPDE1} with $u_{\textrm{b}}=0$ and $(u_0,h,D,v)^\mathsf{T} = (1, 50, 10, 1)^\mathsf{T}$ at $t=50$, where the solution at $x=-200, -195, -190, \ldots, 200$ is corrupted with multiplicative log-normal noise with $\sigma = 0.2$.  Here we see that the data is noise-free as $u \to 0^+$ and the variability is largest near $x=50$ where $u(x,t)$ is a maximum at this time.  

For the multiplicative log-normal noise model framework we have a log-likelihood function of the form
\begin{equation}\label{eq:LogNormalLikelihoodADE}
	\ell(\vec{\theta} \mid u^{\rm{obs}}(x_i,t)) = \sum_{i=1}^{I} \log \left[\phi\left(u^{\rm{obs}}(x_i,t); \log(u(x_i,t)), \sigma^2 \right)\right],
\end{equation}
where  $\phi(x;\mu,\sigma^2)$ is the probability density function of the log-normal$(\mu,\sigma^2)$ distribution. Maximizing this log-likelihood function gives $\hat{\vec{\theta}} =  (0.977,51.776,9.813,0.996)^\mathsf{T}$, and superimposing $u(x,t)$ evaluated with $\hat{\vec{\theta}}$ in Figure \ref{fig:F6}(a) indicates that the solution matches the data reasonably well.  

Given our log-likelihood function and our estimate of $\hat{\vec{\theta}}$, we can work with a normalized log-likelihood function and repeat the construction of the four univariate profile likelihood functions in exactly the same way as we did for the additive Gaussian noise model.  The univariate profile likelihood functions in Figure \ref{fig:F5}(b)--(e) indicate that the four parameters are practically identifiable.  As before, we can sample the log-likelihood function to obtain $M$ samples of $\vec{\theta}$ with $\bar{\ell} \ge \bar{\ell}^{*}$ and use these $M$ solutions of Equation~\eqref{eq:ADEPDE1} to construct the 95\% prediction interval given in Figure \ref{fig:F5}(a).  For the multiplicative noise model we see that the prediction intervals have the attractive property that the upper and lower bounds are always non-negative.  This is very different to working with an additive Gaussian noise model since this procedure can lead to negative prediction intervals which is not physically realistic.

\newpage 
\section{Modeling with a BVP: Dealing with non-identifiability} \label{sec:BVP}

In this section we will work with a linear BVP that  illustrates how non-identifiability can arise, even in very simple mathematical models.  We consider a simple reaction-diffusion equation that has often been used as a caricature model of the morphogen gradients that arise during embryonic development and are thought to be associated with spatial patterning during morphogenesis~\cite{Kicheva2007,Lander2002}
\begin{equation}\label{eq:MorphogenPDE}
\dfrac{\partial u}{\partial t} =  D \dfrac{\partial^2 u}{\partial x^2} - ku, \quad \textrm{on} \quad x > 0,
\end{equation}
where $u(x,t) \ge 0$ represents a non-dimensional morphogen concentration at location $x$ at time $t$.  This model is often considered with the trivial initial condition $u(x,0)=0$.  The morphogen gradient is formed along the $x$-axis by applying a constant diffusive flux in the positive $x$-direction at the origin, giving $J = -D \partial u/ \partial x$ at $x=0$. This simple model assumes that the morphogens undergo diffusion with diffusivity $D > 0$  [L$^2$/T], as well as undergoing some decay process that is modeled with a first-order decay term with decay rate $k > 0$ [/T]. Again, we do not work with specific units for $D$ or $k$, instead we use standard notation~\cite{Barenblatt2003} to indicate that the diffusion coefficient has dimensions of length squared per time, and the decay coefficient has dimensions of per time.  This mathematical model is closed by assuming that the solution vanishes in the far field, that is $u \to 0^+$ as $x \to \infty$.  

\begin{figure}[htp]
  \centering
  \includegraphics[width=\textwidth]{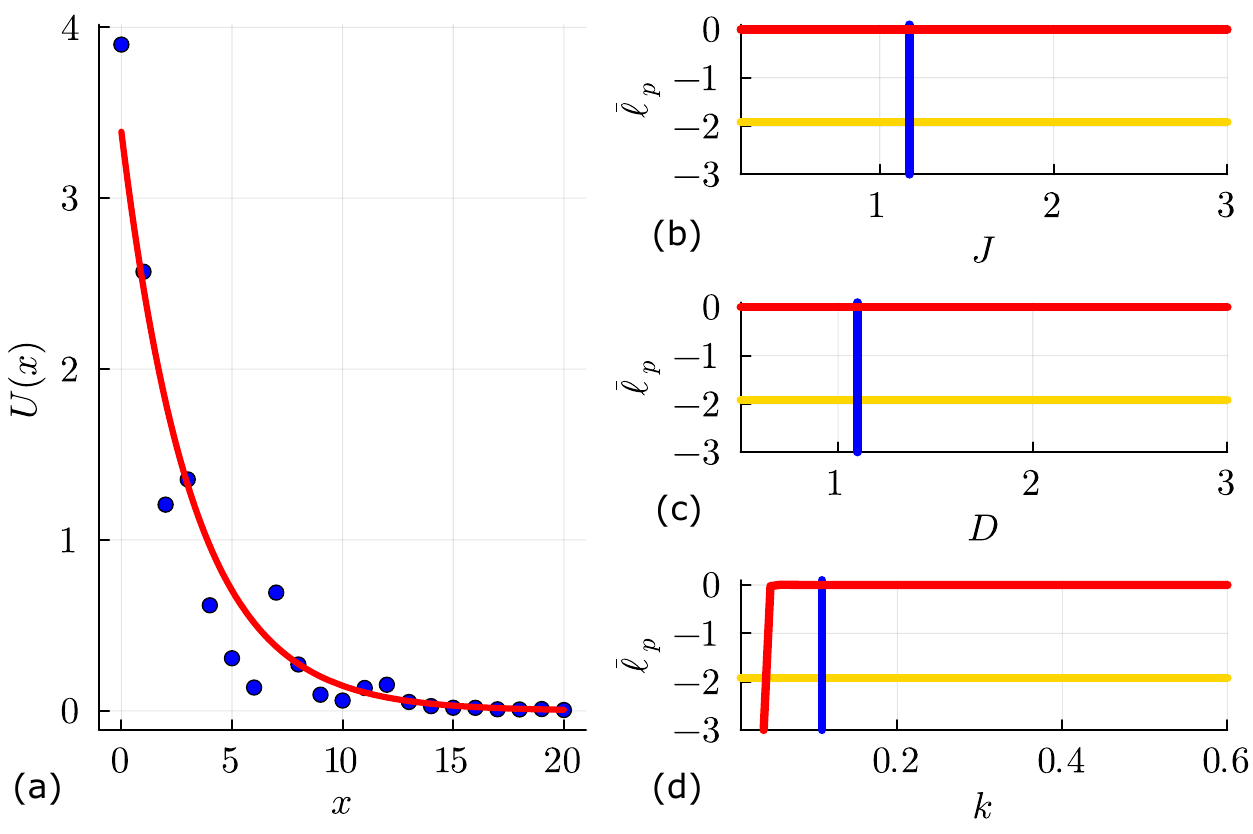}
  \caption{(a) Synthetic data (blue dots) showing observations $U^{\rm{obs}}(x)$ at $x = 0, 2, 4, \ldots, 20$ superimposed with the MLE solution (solid red) with $\hat{\vec{\theta}} = (\hat{J},\hat{D},\hat{k})^\mathsf{T} = (1.171,1.100,0.108)^\mathsf{T}$. (b)--(d) Profile likelihood functions for $J$, $D$ and $k$, as indicated (solid red).  Each profile indicates the MLE (solid blue) and the $95\%$ threshold $\bar{\ell}^{*} = -\Delta_{0.95,1}/2 =  -1.921$ (solid gold). The data are obtained by solving  Equation~\eqref{eq:MorphogenBVP} with $\hat{\vec{\theta}} =  (1,1,0.1)$ and corrupting the solution at  $x = 0, 2, 4, \ldots, 20$ with multiplicative log-normal noise with $\sigma=0.5$. \label{fig:F6}}
\end{figure}

While, in principal, it is possible to use an integral transform to solve Equation~\eqref{eq:MorphogenPDE} to give an expression for $u(x,t)$ like we did in Section \ref{sec:PDE} for the advection-diffusion equation, it is both mathematically convenient and biologically relevant to consider the long-time limit of the time-dependent solution by studying the steady-state distribution, $\displaystyle{\lim_{t \to \infty}}u(x,t) = U(x)$, where $U(x)$ is governed by the following BVP,
\begin{equation}\label{eq:MorphogenBVP}
0 =  D \dfrac{\textrm{d}^2U}{\textrm{d}x^2} - kU, \quad \textrm{on} \quad x > 0,
\end{equation}
with $\textrm{d}U/\textrm{d}x = -J/D$ at $x=0$, and $U \to 0^+$ as $x \to \infty$.  The solution of the steady state BVP can be written
\begin{equation}\label{eq:MorphogenBVPSolution}
U(x) = \dfrac{J}{\sqrt{Dk}}\textrm{exp}\left(-x\sqrt{\dfrac{k}{D}}\right).
\end{equation}
As for the PDE model in Section \ref{sec:PDE}, here we have $U > 0$ by definition.  Accordingly, we present data in Figure \ref{fig:F6}(a) corresponding to $\vec{\theta} = (J,D,k)^\mathsf{T} = (1,1,0.1)^\mathsf{T}$ on the truncated domain $0 < x < 20$.  The solution at $x=0,2,4,\ldots,20$ is corrupted with multiplicative log-normal noise with $\sigma = 0.5$ and, as expected, we see the fluctuations in the data vanish as $x \to \infty$  where $U \to 0^+$.   With this framework we have a log-likelihood function of the form
\begin{equation}\label{eq:LogNormalLikelihoodBVP}
	\ell(\vec{\theta} \mid U^{\rm{obs}}(x_i)) = \sum_{i=1}^{I} \log \left[\phi\left(U^{\rm{obs}}(x_i); \log(U(x_i)), \sigma^2 \right)\right],
\end{equation}
where $\phi(x;\mu,\sigma^2)$ is the probability density function of the log-normal$(\mu,\sigma^2)$ distribution.  An informed approach for parameter estimation with this model and data would first consider a structural identifiability analysis of the mathematical model.  In this instance, for pedagogical reasons, we proceed naively by attempting to estimate parameters without explicitly considering structural identifiability in the first instance.  We will return to discuss structural identifiability of the model later.

To proceed, Numerical optimization gives $\hat{\vec{\theta}} = (\hat{J},\hat{D},\hat{k})^\mathsf{T} = (1.171,1.100,0.108)^\mathsf{T}$, and superimposing $U(x)$ evaluated at $\hat{\vec{\theta}}$ on the data indicates that the solution provides a good match to the data, but as with all previous problems the MLE point estimate provides no insight into parameter identifiability.   The identifiability can be assessed using the exact same procedures implemented in Section \ref{sec:PDE} to give the univariate profiles for $J$, $D$ and $k$ in Figure \ref{fig:F6}(b)--(d).  These univariate profile likelihood functions immediately indicate that these parameters are not well identified by this data because the profiles are flat.  These flat profiles indicate that there are many different parameter choices for which Equation~\eqref{eq:MorphogenBVPSolution} matches the data equally well, which is an example of structural non-identifiability in this case.

In this example the structure of the model solution, Equation~\eqref{eq:MorphogenBVPSolution}, indicates that the parameters are structurally non-identifiable because $U(x)$ depends only upon two particular parameter combinations, namely $J/\sqrt{kD}$ and $\sqrt{k/D}$.  Since there are infinitely many choices of $J$, $D$ and $k$ that give the same values for $J/\sqrt{kD}$ and $\sqrt{k/D}$, we expect the profile likelihood functions in Figure \ref{fig:F6}(b)--(d) to be flat.

\begin{figure}[htp]
  \centering
  \includegraphics[width=\textwidth]{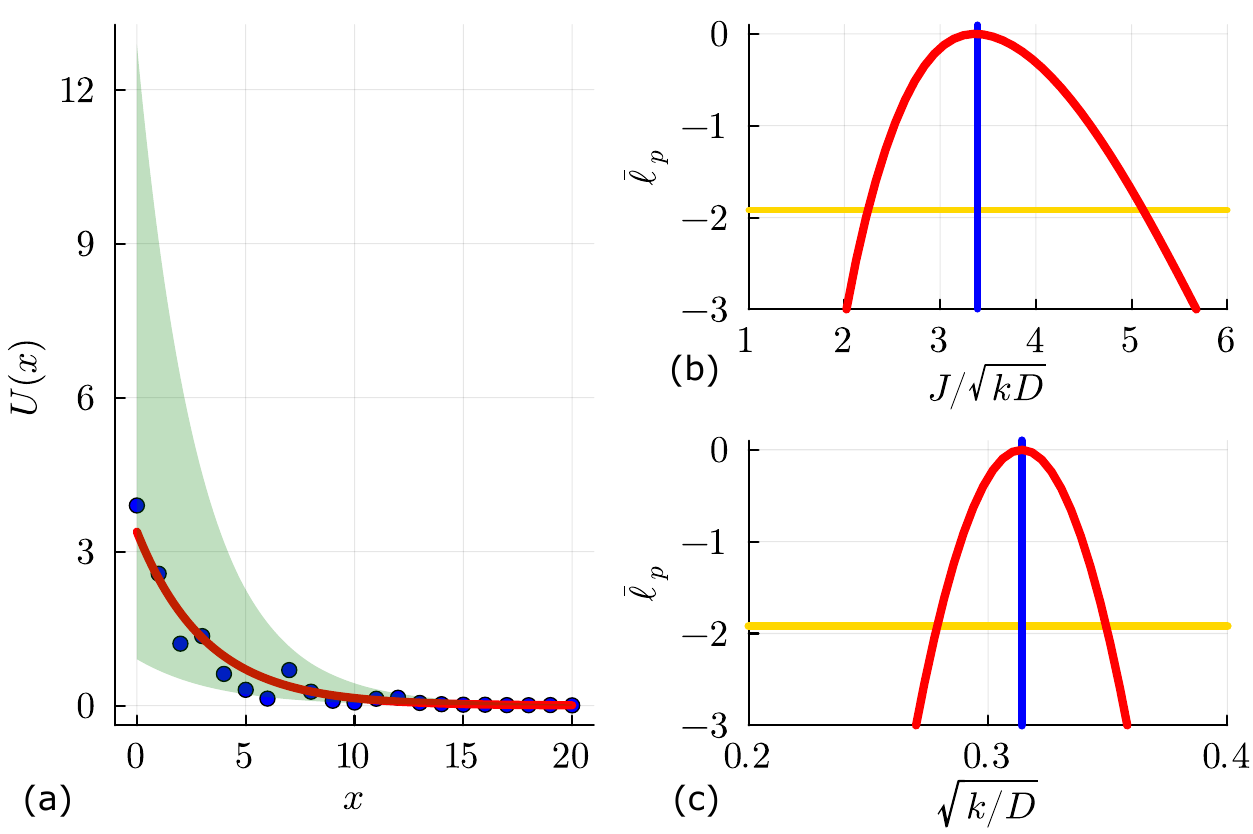}
  \caption{(a) Synthetic data (blue dots) showing observations $U^{\rm{obs}}(x)$ at $x = 0, 2, 4, \ldots, 20$ superimposed with the MLE solution (solid red) with $\hat{\vec{\theta}}_\text{r} = (\hat{\alpha},\hat{\beta})^\mathsf{T} = (3.389, 0.3141)^\mathsf{T}$ and the 95\% prediction interval (shaded green region). (b)--(c) Profile likelihood functions for $\alpha = J/\sqrt{kD}$ and $\beta = \sqrt{k/D}$, as indicated (solid red).  Each profile indicates the MLE (solid blue) and the $95\%$ threshold $\hat{\ell}^{*} = -\Delta_{0.95,1}/2 =  -1.921$ (solid gold). The 95\% confidence intervals are $\alpha = J/\sqrt{kD}\in[2.243,5.121]$, $\beta = \sqrt{k/D}\in[0.279, 0.349]$. The data are obtained by solving Equation~\eqref{eq:MorphogenBVP} with $\hat{\vec{\theta}} =  (1,1,0.1)$ and corrupting the solution at  $x = 0, 2, 4, \ldots, 20$ with multiplicative log-normal noise with $\sigma=0.5$.\label{fig:F7}}
\end{figure}

In this problem the structure of the exact solution suggests a re-parameterisation of the likelihood function to remove the identifiability issue. We take $\vec{\theta}_\text{r} = (\alpha,\beta)^\mathsf{T}$, where $\alpha = J/\sqrt{kD}$ and $\beta = \sqrt{k/D}$, and attempt to estimate $\vec{\theta}_\text{r} = (\alpha,\beta)^\mathsf{T}$ instead of $\vec{\theta} = (J, D, k)^\mathsf{T}$.   Results in Figure \ref{fig:F7}(a)--(c) re-examine the same data using the re-parameterized log-likelihood function, and numerical optimization gives $\hat{\vec{\theta}}_\text{r}=(3.389,0.3141)^\mathsf{T}$.  We examine the identifiability of the re-scaled parameters by constructing univariate profile likelihood functions for $\alpha = J/\sqrt{kD}$ and $\beta = \sqrt{k/D}$ and we find that both quantities are well identified by the data.  Repeating the process of sampling $M$ values of $\vec{\theta}_\text{r} = (\alpha,\beta)^\mathsf{T}$ where $\bar{\ell} \ge \bar{\ell}^{*}$ and using these $M$ solutions of Equation~\eqref{eq:MorphogenBVP} to construct the 95\% prediction interval given in Figure \ref{fig:F7}(a) where we again see that the prediction intervals have the useful property that they are non-negative.

\section{Extensions and general remarks} \label{sec:conclusion}
This article describes a set of self-contained computational exercises that develop awareness, knowledge and skills relating to parameter estimation, parameter identifiability and model prediction.  A key aim of these exercises is to illustrate how a likelihood-based approach can be applied to a range of different process models that are of interest to applied mathematicians (e.g.~ODE-, PDE-, BVP-based models). Details of all computational exercises can be repeated and extended using open source software available on \href{https://github.com/ProfMJSimpson/IdentifiabilityTutorial}{GitHub}.

The example problems dealt within this article reflect a compromise between keeping all calculations sufficiently straightforward while also working with mathematical modeling scenarios of practical interest.  Accordingly there are many ways that the examples can be extended.  For example, all data considered in this article are generated using either an additive Gaussian noise model or a multiplicative log-normal noise model with a known variance.  While sometimes it is possible to pre-estimate the value of $\sigma^2$ from a real data set~\cite{Simpson2020}, it is also straightforward to extend the vector of unknown parameters and treat $\sigma$ as an unknown quantity to be determined along with the other model parameters~\cite{Simpson2022}.  Similarly, when we dealt with the PDE model in Section \ref{sec:PDE} we considered data at several spatial locations, $u^{\rm{obs}}(x_i,t)$ for $i=1,2,3,\ldots, I$, but just one fixed point in time.  In many situations data are available at different points in space and at different times, $u^{\textrm{o}}(x_i,t_j)$ for $i=1,2,3,\ldots, I$ and $j=1,2,3,\ldots, J$, and dealing with such data is straightforward by summing over all data points, both in space and time, in the log-likelihood function, given by Equation~\eqref{eq:LogNormalLikelihoodADE}.  

Another feature of the examples presented in this article is that all process models are analytically tractable differential equations.  In general, analytically tractable models are usually limited to special cases, such as dealing with linear differential equations.  Therefore, another insightful extension is to replace the use of the various exact solutions with a numerical solution, obtained for example using the DifferentialEquations.jl package in Julia to solve time-dependent ODE models~\cite{Rackauckas2017}.  Repeating the exercises in this article using numerical solutions will be a useful stepping-stone for readers who are interested in using more general process models based on nonlinear differential equations where exact solutions are not always possible. 

We assess parameter identifiability by using numerical optimization to construct various profile likelihood functions.  Other approaches are possible, such as calculating the Hessian at the MLE, and assuming $\bar{\ell}$ can be approximated by a low order truncated Taylor series~\cite{Wasserman2004}. Such local approaches, based on information at the $\hat{\vec{\theta}}$ can be insightful for identifiable problems.  Unfortunately, such local approaches can give misleading results when applied to non-identifiable problems.  From this point of view, the profile likelihood can be viewed as a global approach by accounting for properties of $\bar{\ell}$ away from $\hat{\vec{\theta}}$.

In terms of making model predictions, here we always use a very simple rejection sampling method to find $M$ samples of $\vec{\theta}$ where $\bar{\ell} \ge \bar{\ell}^{*}$. We chose to use rejection sampling because it is both simple to implement and interpret, but other approaches are possible.  For example, we could have simply evaluated $\bar{\ell}$ across a  uniform grid to propagate the parameter confidence set through to examine model predictions.  Both approaches carry advantages and disadvantages and give very similar results provided that $M$ is chosen to be sufficiently large when using rejection sampling, and that the uniform mesh is taken to be sufficiently dense when working with a gridded log-likelihood function.

Our results in Section \ref{sec:BVP} demonstrate how non-identifiability manifests as flat univariate profile likelihood functions, and in this  case we are able to use the exact solution of the BVP to motivate a simple re-parameterization of the log-likelihood function.  This approach, while instructive, is not always possible when the process model is intractable.  In such cases it is sometimes possible to use different methods to determine appropriate re-parameterization options~\cite{Cole2020,Simpson2022}.

A key feature of the computational exercises in this work is that we used synthetic data collected from a particular model, and then we used the same model for parameter estimation.  This is convenient to demonstrate and explore computational techniques, however in real-world applications there is always some inherent uncertainty in the mechanisms acting to produce experimental data.  As such, the question of model selection arises because there is almost always more than one possible process model that can replicate and interrogate the data.  In these cases the tools developed here to explore parameter identifiability, parameter estimation and model prediction can be used across a number of competing models to help in the process of model selection.  In this situation it can be useful to compute profile likelihood functions for each parameter in the competing models to help to rule out working with a model involving non-identifiable parameters when alternative identifiable models are available.  Similarly, when dealing with experimental data it can be useful to compute and compare profile likelihood functions using different noise models to guide the selection of an appropriate noise model.  A final comment is that all of the exercises presented in this work deal with deterministic mathematical models.  It is possible to apply the same ideas to stochastic mathematical models by using, for example, a coarse-grained continuum limit description of the stochastic model~\cite{Codling2008,Plank2025,Simpson2021}.

\printbibliography

\end{document}